# Verification Method for Graph Isomorphism Criteria

Hu Chuanfu[1], Hou Aimin[2]

[1]Dongguan University of Technology, [2]Dongguan City University

[1]hucf@dgut.edu.cn, [2]houam@dgut.edu.cn

## Abstract

The criteria for determining graph isomorphism are crucial for solving graph isomorphism problems. The necessary condition is that two isomorphic graphs possess invariants, but their function can only be used to filtrate and subdivide candidate spaces. The sufficient conditions are used to rebuild the isomorphic reconstruction of special graphs, but their drawback is that the isomorphic functions of subgraphs may not form part of the isomorphic functions of the parent graph. The use of sufficient or necessary conditions generally results in backtracking to ensure the correctness of the decision algorithm. The sufficient and necessary conditions can ensure that the determination of graph isomorphism does not require backtracking, but the correctness of its proof process is difficult to guarantee. This article proposes a verification method that can correctly determine whether the judgment conditions proposed by previous researchers are sufficient and necessary conditions. A subdivision method has also been proposed in this article, which can obtain more subdivisions for necessary conditions and effectively reduce the size of backtracking space.

## 1  Introduction

The isomorphism determination of graphs has wide applications in fields such as computational complexity theory, pattern recognition, computer vision processing, data mining, VLSI design validation, and chemical molecular structure recognition[1-3]. In the 1970s, Karp, Garey, and Johnson proved that the isomorphism determination problem of graphs was one of the few particular problems that could neither be classified as **P** nor **NP**-hard[1]. Since then, this problem has become a public issue, which occupies a special position, in the field of theoretical computers. On the one hand, graph isomorphism problems have not yet been proven to be **NPC** problems, so people have great aspirations to find polynomial decision algorithms. On the other hand, people also hope to prove that there is no polynomial decision algorithm for graph isomorphism problems, thereby proving that there is indeed another class of problems between the **P** class and the **NPC** class; This indirectly proves the conjecture that **P≠NP**.

Given two graphs $G$ and $H$, the vertex sets are $V(G)$ and $V(H)$, respectively. If there exists a bijective function $f: V(G) \to V(H)$ such that $(u, v)$ is an edge of $G$ if and only if $(f(u), f(v))$ is an edge of $H$, then $G$ and $H$ are isomorphic. When a bijective function $f$ does not satisfy this definition, it is necessary to find other bijective function $f'$. Unfortunately, there are n! bijective functions, where $n=|V(G)|$. When it is verified that n! bijective functions do not satisfy the isomorphism definition, it can be determined that graphs $G$ and $H$ are not isomorphic. The various decision algorithms proposed so far aim to reduce the size of the search space through various pruning strategies. The known optimal search space size is $\exp((\log n)^{o(1)})$.

Two isomorphic graphs have many similar properties[4], such as the same degree sequence, the same adjacency field, the same maximum spanning subtree, the same characteristic polynomial, the same radius, the same eccentricity, the same distance spectrum, and so on. Therefore, necessary conditions, sufficient conditions, sufficient and necessary conditions, and judgment algorithms have always been highly concerned and many high-level research results about them have been achieved. Firstly, the necessary condition is the essential property of two isomorphic graphs, called invariants. As a filtrating technique, it is applied in decision algorithms to narrow down the search space range of candidate isomorphic vertices. Secondly, sufficient conditions apply to the reconstruction of special graphs or the reconstruction of special parameters of general graphs, and cannot be applied to the reconstruction of general graphs. The Ulam conjecture is a measure of the reconstruction of general graphs, but it has not been proven to date. There are some articles in literatures that claim to have found "sufficient and necessary conditions", but later it has been proven that they are not sufficient and necessary conditions. Recently, Wang Zhuo and Wang Chenghong[5] used distance matrices to calculate characteristic polynomials as a necessary and sufficient condition, and published their results in the ACTA AUTOMATICA SINICA (in Chinese). Their proof is based solely on the relevant theories of linear algebra and matrix theory, and cannot distinguish its essential difference from the proof process that uses adjacency matrices to calculate characteristic polynomials as sufficient and necessary conditions. Unfortunately, the characteristic polynomials of adjacency matrix has been proven through counterexamples that it is not a sufficient and necessary condition.

Undoubtedly, counterexamples can prove that a so-called "sufficient and necessary condition" is wrong. But designing very "clever" counterexamples is very difficult. Another approach is to use the existing datasets to prove it being wrong or untrue, but there are cases where the dataset cannot give the refutation. This article proposes a "very simple" verification method that can determine whether there is backtracking in the process of finding isomorphic functions based on any two isomorphic graphs. If there is backtracking, we can prove that the so-called "sufficient and necessary conditions" are incorrect conclusions. Further, we can design a judgment algorithm for finding isomorphic functions based on this verification method.

In addition to the decision conditions, designing judgment algorithms is also one of the research hotspots. In 1982, Luks[6] used mathematical tools such as finite groups to provide the best algorithm for determining isomorphism between two graphs at that time. The time complexity of his algorithm was $\exp(O(\sqrt{n}\log n))$ (where *n* is the number of vertices in the graph). In the following 40 years, the problem of isomorphism determination of graphs has attracted the attention of many scholars, and hundreds of articles in this area have been published in different academic journals[1]. 1) Isomorphism determination algorithms (such as Ullman algorithm, Schmidt algorithm, Falkenhainer algorithm, Messmer algorithm, etc.) for some special graphs (such as trees, bounded interval graphs, maximal outer plane graphs, etc.)[6, 8-9] were proposed. These algorithms mainly use "vertex normalized labeling or vertex partition division" strategy and "backtracking searching" technique, and have been proven to have polynomial time complexity. 2) Conventional judgment algorithms for general simple graphs, such as some programs on the Internet used to test whether two graphs are isomorphic (such as Nauty, Saucy, Bliss, Conauto, and Traces, etc.)[3] were proposed. These programs run very fast, but the time complexity analysis and correctness proof of the algorithms have not been publicly reported. 3) The decision algorithms for intelligent computing of general simple graphs, such as genetic algorithms, neural networks, particle swarm optimization algorithms, DNA algorithms, and quantum computing algorithms[4, 10] were proposed. Although they are efficient, their time complexity is difficult to accurately analysis. In 2015, Babai[7] proposed a quasi-polynomial algorithm for the isomorphism determination problem of two graphs based on the Luks's algorithm, utilizing the local relationships between orbits under group action and the regular decomposition technique of groups, with the time complexity $\exp((logn)^{o(1)})$.

The main contributions of this work include: 1) We propose a simple and effective verification method that can correctly determine whether a judgment condition is sufficient and necessary condition or not without the need to use counterexamples. 2) Based on the principle of reconstruction of isomorphic graphs, a more efficient subdivision method for vertex set partitioning was designed using the "induced subgraph + remaining subgraph" techniques to find isomorphic functions, and better performance results were obtained than existing algorithms.

In Section 2, the preparatory knowledge required for this article is introduced. The verification method and the subdivision method proposed in this article are discussed in Section 3 and Section 4 respectively. In Section 5, the correctness and effectiveness of the proposed methods are verified through case analysis. Finally, a summary of our conclusions is presented in Section 6.

## 2  Preliminaries

**Definition 1.** Let $G_1=<V_1, E_1>$ and $G_2=<V_2, E_2>$ be two simple graphs, $|V_1|=|V_2|=n\geq 2$. If there exists a bijective function $f: V_1 \to V_2$ such that $(u, v)$ is an edge of $G_1$ if and only if $(f(u), f(v))$ is an edge of $G_2$, then $G_1$ and $G_2$ are isomorphic.

**Lemma 1**. Let $G_1=<V_1, E_1>$ and $G_2=<V_2, E_2>$ be two simple graphs, where $\zeta$ is a criterion. $\zeta$ is a necessary condition if and only if $G_1 \cong G_2 \Rightarrow \zeta(G_1)=\zeta(G_2)$.

**Corollary 1**. Let $G_1=<V_1, E_1>$ and $G_2=<V_2, E_2>$ be two simple graphs, where $\zeta$ is a criterion. If $\zeta(G_1)\neq\zeta(G_2) \Rightarrow G_1 \not\cong G_2$, then $\zeta$ is a necessary condition.

**Lemma 2.** Let $G_1=<V_1, E_1>$ and $G_2=<V_2, E_2>$ be two simple graphs, where $\zeta$ is a criterion. $\zeta$ is a sufficient condition if and only if $\zeta(G_1)=\zeta(G_2) \Rightarrow G_1 \cong G_2$.

**Corollary 2**. Let $G_1=<V_1, E_1>$ and $G_2=<V_2, E_2>$ be two simple graphs, where $\zeta$ is a criterion. If $G_1 \not\cong G_2 \Rightarrow \zeta(G_1)\neq\zeta(G_2)$, then $\zeta$ is a sufficient condition.

**Lemma 3.** Let $G_1=<V_1, E_1>$ and $G_2=<V_2, E_2>$ be two simple graphs, where $\zeta$ is a criterion. $\zeta$ is a sufficient and necessary condition if and only if $G_1 \cong G_2 \Leftrightarrow \zeta(G_1)=\zeta(G_2)$.

Proof. Since $G_1 \cong G_2 \Leftrightarrow \zeta(G_1)=\zeta(G_2)$, so $G_1 \cong G_2 \Rightarrow \zeta(G_1)=\zeta(G_2)$ and $\zeta(G_1)=\zeta(G_2) \Rightarrow G_1 \cong G_2$. We can infer that $\zeta$ is a necessary condition due to $G_1 \cong G_2 \Rightarrow \zeta(G_1)=\zeta(G_2)$. And we can infer that $\zeta$ is a sufficient condition due to $\zeta(G_1)=\zeta(G_2) \Rightarrow G_1 \cong G_2$.

Therefore, $\zeta$ is a sufficient and necessary condition. □

**Definition 2.** Let $G_1=<V_1, E_1>$ and $G_2=<V_2, E_2>$ be two simple graphs and $G_1 \cong G_2$. If there are two vertices $u$ and $v$ such that $G_1+u\cong G_2+v$, then $G_1+u$ and $G_2+v$ are said to be reconfigurable.

According to **Definition 2**, if $G_1+u$ and $G_2+v$ are reconfigurable, there must exist an isomorphic function $f$ between $G_1$ and $G_2$, such that $v=f(u)$, and $\forall w\in V_1$, $(w, u)\in E_1 \Leftrightarrow (f(w), v)\in E_2$.

**Lemma 4.** Let $G_1=<V_1, E_1>$ and $G_2=<V_2, E_2>$ be two simple graphs, and $f$ is a isomorphism function of $G_1 \cong G_2$. If $G_1+u$ and $G_2+v$ are reconfigurable, then $N(v)=f(N(u))$. Here, $N(u)$ represents the adjacency field of vertex $u$.

**Corollary 3**. Let $G_1=<V_1, E_1>$ and $G_2=<V_2, E_2>$ be two simple graphs, and $f$ is a isomorphism function of $G_1 \cong G_2$. If $N(v) \neq f(N(u))$, then $u$ and $v$ are not the reconstruction corresponding points of $G_1+u$ and $G_2+v$ under the action of $f$.

**Theorem 1** (Verification Method)**.** Let $G_1 = <V_1, E_1>$ and $G_2 = <V_2, E_2>$ be two simple graphs, and $\zeta$ is a sufficient and necessary condition. So, $\zeta(G_1) = \zeta(G_2) \Rightarrow G_1 \cong G_2$. Assuming the isomorphic function is $f$. For $u \in V_1$, $\exists v \in V_2$, such that $G_1 - u \cong G_2 - v$, $v = f(u)$. The isomorphic function $f$ still maintains the isomorphism relationship between $G_1$-$u \cong G_2$-$v$. The property that the generated subgraph obtained by deleting some vertex is isomorphic and $f$ still maintains the isomorphic relationship remains true until the generated subgraph is empty (i.e. there are no vertices).

Proof. Since $\zeta$ is a sufficient and necessary condition, we have $\zeta(G_1^{(t)}) = \zeta(G_2^{(t)}) \Rightarrow G_1^{(t)} \cong G_2^{(t)}$ for any two simple graphs $G_1^{(t)}$ and $G_2^{(t)}$. Assuming $G_1 \cong G_2$ and the isomorphic function is $f$. For $\forall u, u' \in V_1$, there exist $v = f(u)$, $v' = f(u') \in V_2$, such that $(u, u') \in E_1 \Leftrightarrow (v, v') \in E_2$, where $f = \{(u, v) \mid v = f(u)\}$.

Select any element $(u, v)$ from $f$, let $G_1^{(1)} = G_1 - u$, $G_2^{(1)} = G_2 - v$. So we have $\forall u', u'' \in V_1$, $v' = f(u') \in V_2$, $v'' = f(u'') \in V_2$, such that $(u, u') \in E_1 \Leftrightarrow (v, v') \in E_2$, $(u, u'') \in E_1 \Leftrightarrow (v, v'') \in E_2$, $(u', u'') \in E_1 \Leftrightarrow (v', v'') \in E_2$. Therefore, $G_1^{(1)} \cong G_2^{(1)}$, the new isomorphic function is $f' = f - \{(u, v)\}$. In other words, under the action of $f'$, $G_1^{(1)} + u \cong G_2^{(1)} + v$ can be reconstructed by $G_1^{(1)} \cong G_2^{(1)}$. Here, $u$ and $v$ are reconstruction corresponding points, and the isomorphic function $f = f' + \{(u, v)\}$.

Continuously, select any element $(u', v')$ from $f'$, let $G_1^{(2)} = G_1^{(1)} - u'$, $G_2^{(2)} = G_2^{(1)} - v'$. The above proof process still holds true. And so forth, finally there is only one element $(u^{(n-1)}, v^{(n-1)})$ in $f^{(n-1)}$. Let $G_1^{(n)} = G_1^{(n-1)} - u^{(n-1)}$, $G_2^{(n)} = G_2^{(n-1)} - v^{(n-1)}$. Then $G_1^{(n)}$ and $G_2^{(n)}$ are empty graphs. □

Note that in the iterative process of generating subgraphs for reconfigurability in **Theorem 1**, there is no backtracking operation. This is very important as it forms the cornerstone of our verification method. In other words, use $\zeta(G_1) = \zeta(G_2) \Rightarrow G_1 \cong G_2$ to determine the isomorphism corresponding points $u$ and $v$. Then use $\zeta(G_1 - u) = \zeta(G_2 - v) \Rightarrow G_1 - u \cong G_2 - v$ to determine the next isomorphism corresponding points $u'$ and $v'$ until the isomorphism function $f$ of $G_1 \cong G_2$ is obtained. There is no backtracking in this iterative process. On the contrary, if backtracking is unavoidable, then the judgment condition $\zeta$ is not a sufficient or necessary condition.

Obviously, the verification method determined by **Theorem 1** is easier to implement than finding counterexamples to deny the necessary and sufficient condition.

**Definition 3.** Let $A$ be a non empty set, $S = \{S_1, S_2, \ldots, S_k\}$. If $S$ satisfies the following conditions: (1) $S_i \neq \Phi$; (2) $S_i \subseteq A$; (3) $S_i \cap S_j = \Phi$; (4) $\bigcup_{i=1}^{k} S_i = A$. Then set $S$ is called a partition of $A$, and $S_i$ is a partition block or partition unit of $A$.

**Definition 4.** Given any two partitions $\{A_1, A_2, \ldots, A_r\}$ and $\{B_1, B_2, \ldots, B_s\}$ of set $X$. If for each $A_i$, there is a $B_j$ such that $A_i \subseteq B_j$. Then $\{A_1, A_2, \ldots, A_r\}$ is a subdivision of $\{B_1, B_2, \ldots, B_s\}$.

**Theorem 2**. The partition determined by the degree sequence of adjacent vertices is a subdivision of the partition determined by vertex degrees.

Proof. Assuming that the partition determined by the degree sequence of adjacent vertices is $\{A_1, A_2, \ldots, A_r\}$, and the partition determined by vertex degrees is $\{B_1, B_2, \ldots, B_s\}$.

Situation 1: If we have $A_i = B_i$, $r = s$, $1 \leq i \leq r$, then $\{A_1, A_2, \ldots, A_r\} = \{B_1, B_2, \ldots, B_s\}$.

Situation 2: For any two vertices $u$ and $v$, if they are not in the same partition block $B_i$, there must be deg($u$)≠deg($v$), inferring that the length of the adjacency degree sequence of vertex $u$ ≠ the length of the adjacency degree sequence of vertex $v$. So vertices $u$ and $v$ do not have the same degree sequence of adjacent points. Therefore, vertices $u$ and $v$ are not in the same partition block $A_j$.

Situation 3: For any two vertices $u$ and $v$, if they are in the same partition block $B_i$, there must be deg($u$)=deg($v$).

Scenario 3.1. If the degree sequence of adjacent points of vertex $u$ = the degree sequence of adjacent points of vertex $v$, then vertices $u$ and $v$ are in the same partition block $A_j$.

Scenario 3.2. If the degree sequence of adjacent points of vertex $u$ ≠ the degree sequence of adjacent points of vertex $v$, then vertices $u$ and $v$ are not in the same partition block $A_j$.

On the other hand, for any pair of vertices $u$ and $v$ in $A_j$, we must have the degree sequence of adjacent points of vertex $u$ = the degree sequence of adjacent points of vertex $v$. So deg($u$)=deg($v$), inferring that $u$ and $v$ are in the same $B_i$.

Taking into account all the above situations, for each $A_j$, there is a $B_i$, such that $A_j \subseteq B_i$.  □

The degree sequence of adjacent vertices and the degree of vertices are both necessary conditions for graph isomorphism. **Theorem 2** states that there exists a necessary condition, and the partition determined by it is always a subdivision of the partition determined by another necessary condition.

**Theorem 3** (Subdivision Method). For any given criterion $\zeta$, we can construct an induced subgraph $N(u)+u$ as well as a remaining subgraph $G-(N(u)+u)$ derived from the vertex $u$ which is determined by $\zeta(G-u)$. The division determined by the latter is a subdivision of the division determined by the former.

Proof. Assuming that the partition determined by a certain judgment condition $\zeta$ is {$A_1$, $A_2$, …, $A_r$}, the partition obtained by applying the "induced subgraph + remaining subgraph" strategy determined by $N(u)+u$ is {$B_1$, $B_2$, …, $B_s$}.

Situation 1: If we have $A_i=B_i$, $r=s$, $1 \leq i \leq r$, then {$A_1$, $A_2$, …, $A_r$}={$B_1$, $B_2$, …, $B_s$}.

Situation 2: For any two vertices $u$ and $v$, if they are not in the same partition block $A_i$, there must be $\zeta(G-u) \neq \zeta(G-v)$, inferring that the induced subgraph $N(u)+u$ ≠ the induced subgraph $N(v)+v$; or, the remaining subgraph $G-(N(u)+u)$ ≠ the remaining subgraph $G-(N(v)+v)$. Therefore, vertices $u$ and $v$ are not in the same partition block $B_j$.

Situation 3: For any two vertices $u$ and $v$ in any partition block $A_i$ ($1 \leq i \leq r$), they are either the real "isomorphic vertices" or not the real "isomorphic vertices". If the induced subgraph as well as the remaining subgraph derived from the set of $N(u)+u$ and the set of $N(v)+v$ respectively are the same, the vertices $u$ and $v$ are all in the same $B_j$ ($1 \leq j \leq s$). Otherwise, they are not in the same $B_j$ ($1 \leq j \leq s$).

On the other hand, for any pair of vertices $u$ and $v$ in $B_j$, we must have the induced subgraph $N(u)+u$ = the induced subgraph $N(v)+v$, as well as, the remaining subgraph $G-(N(u)+u)$ = the remaining subgraph $G-(N(v)+v)$. So $\zeta(G-u)=\zeta(G-v)$, inferring that $u$ and $v$ are in the same $A_i$.

So, for each $B_j$, there is a $A_i$, such that $B_j \subseteq A_i$.  □

## 3  Verification Method

Based on **Theorem 1**, this paper proposes a verification method that can correctly determine whether a given judgment condition $\zeta$ is a sufficient and necessary condition within $O(n^2)$ time. On the premise that $\zeta$ is a sufficient and necessary condition and can be calculated in polynomial time, an isomorphic function $f$ can be found in polynomial time. The greatest characteristic of this verification method is that it is independent of the proof process that $\zeta$ is a sufficient and necessary condition. Therefore, it is convenient to identify the authenticity of any sufficient and necessary conditions. Dom($f$) represents the domain of function $f$, Ran($f$) represents the range of function $f$.

**Algorithm:** Verification method for determining the authenticity of sufficient and necessary conditions

**Input:** The adjacency matrices $A_1$ and $A_2$ of two graphs $G_1$ and $G_2$, a judgment condition $\zeta$

**Output:** A isomorphic function $f$, or "two graphs are not isomorphic", or "$\zeta$ is not sufficient and necessary condition"

**Step 1.** Calculate the corresponding $\zeta(A_1)$ and $\zeta(A_2)$ based on $A_1$ and $A_2$;

**Step 2.** If $\zeta(A_1) \neq \zeta(A_2)$, return "two graphs are not isomorphic" and stop the algorithm; Otherwise, go to **Steps 3-12**;

**Step 3.** While ($V(G_1)\neq\{\}$)

**Step 4.** Consider an "unvisited" vertex $u$ of graph $G_1$ and sign it as "visited";

**Step 5.** If all "unvisited" vertices $v$ in graph $G_2$ have been processed, return "$\zeta$ is not a sufficient and necessary condition" and stop the algorithm; Otherwise, go to **Step 6**;

**Step 6.** Consider an "unvisited" vertex $v$ of graph $G_2$ in turn and sign it as "visited";

**Step 7.** Calculate the adjacency matrices $A'_1$ and $A'_2$, as well as the $\zeta(A'_1)$ and $\zeta(A'_2)$ based on the subgraphs $G'_1=G_1$-$u$ and $G'_2=G_2$-$v$, respectively;

**Step 8.** If $\zeta(A'_1) \neq \zeta(A'_2)$, abandon the vertex $v$ and go to **Step 5**, considering another "unvisited" vertex $v$ in graph $G_2$;

**Step 9.** $\forall x \in$Dom($f$), determine whether $(x, u)\Leftrightarrow(f(x), v)$ holds. If it does not hold, abandon the vertex $v$ and go to **Step 5**, considering another "unvisited" vertex $v$ in graph $G_2$; Otherwise, go to **Step 10**;

**Step 10.** Let $G_1=G'_1$, $G_2=G'_2$, add the vertex pair $(u, v)$ to $f$, namely $v=f(u)$, clear up all "visited" tag in $G_2$, go to **Step 3**;

**Step 11.** End of While;

**Step 12.** Return the isomorphic function $f$ and stop the algorithm;

Assuming that the judgment condition $\zeta$ is a necessary and sufficient condition, then under the premise of $\zeta(G_1)=\zeta(G_2)$, graphs $G_1$ and $G_2$ are isomorphic. Therefore, for each vertex $u$ in graph $G_1$, there must be a corresponding point $v$ in graph $G_2$, such that $\forall x \in V(G_1)$, $(x, u)\Leftrightarrow(f(x), v)$. So that, **Steps 8** and **9** are deterministic and do not require backtracking. Contrarily, under the premise of $\zeta(G_1)=\zeta(G_2)$, if there exists a vertex $u$ in graph $G_1$ such there is no corresponding point $v$ in graph $G_2$, that $\forall x \in V(G_1)$, $(x, u)\Leftrightarrow(f(x), v)$. That proves that the judgment condition $\zeta$ is not a sufficient or necessary condition.

Without considering the calculation time cost of the judgment condition $\zeta$, the outer loop (**Step 3**) runs $n$ times, and the inner loop (**Step 5**) runs $n$ times. Therefore, this validation algorithm can be completed in O($n^2$) time.

## 4 Subdivision Method

We have modified the verification method to allow it to backtrack. In order to minimize the number of backtracking, the technique of "induced subgraph + remaining subgraph" is introduced for subdivision of a partition determined by a certain invariant $\zeta$. $N(u)$ represents the adjacency field of vertex $u$, and $N(A)$ represents the union set of the adjacency fields of all vertices in set $A$, Dom($f$) represents the domain of function $f$, Ran($f$) represents the range of function $f$.

**Algorithm:** Subdivision method for subdividing the partition of an invariant using the strategy of "induced subgraph + remaining subgraph"

**Input:** The adjacency matrices $A_1$ and $A_2$ of two graphs $G_1$ and $G_2$, a judgment condition $\zeta$

**Output:** A isomorphic function $f$, or "two graphs are not isomorphic"

**Step 1.** Calculate the corresponding $\zeta(A_1)$ and $\zeta(A_2)$ based on $A_1$ and $A_2$;

**Step 2.** If $\zeta(A_1) \neq \zeta(A_2)$, return "two graphs are not isomorphic" and stop the algorithm; Otherwise, go to **Steps 3-15**;

**Step 3.** While ($V(G_1) \neq \{\}$);

**Step 4.** If all "unvisited" vertices in graph $G_1$ have been processed, but the isomorphic function $f$ cannot be found, then return "two graphs are not isomorphic" and stop the algorithm; Otherwise, go to **Step 5**;

**Step 5.** Consider an "unvisited" vertex $u$ of graph $G_1$ and sign it as "visited";

**Step 6.** If all "unvisited" vertices in graph $G_2$ have been processed, but the isomorphic function $f$ cannot be found, then go to **Step 4**, considering another "unvisited" vertex $u$ in graph $G_1$;

**Step 7.** Consider an "unvisited" vertex $v$ of graph $G_2$ in turn and sign it as "visited";

**Step 8.** Calculate the adjacency matrices $A'_1$ and $A'_2$, as well as the $\zeta(A'_1)$ and $\zeta(A'_2)$ based on the subgraphs $G'_1 = G_1 - u$ and $G'_2 = G_2 - v$, respectively;

**Step 9.** If $\zeta(A'_1) \neq \zeta(A'_2)$, abandon the vertex $v$ and go to **Step 6**, considering another "unvisited" vertex $v$ in graph $G_2$;

**Step 10.** $\forall x \in $Dom($f$), determine whether $(x, u) \Leftrightarrow (f(x), v)$ holds. If it does not hold, abandon the vertex $v$ and go to **Step 6**, considering another "unvisited" vertex $v$ in graph $G_2$; Otherwise, go to **Step 11**;

**Step 11.** Determine whether $\zeta$(the adjacency matrix of $N$(Dom($f$)$\cup\{u\}$))=$\zeta$(the adjacency matrix of $N$(Ran($f$)$\cup\{v\}$)) holds, as well as, $\zeta$(the adjacency matrix of $N(G$-Dom($f$)$\cup\{u\}$))=$\zeta$(the adjacency matrix of $N(G$-Ran($f$)$\cup\{v\}$)) holds. If they do not hold, abandon the vertex $v$ and go to **Step 6**, considering another "unvisited" vertex $v$ in graph $G_2$; Otherwise, go to **Step 12**;

**Step 12.** Determine whether $f$(the induced subgraph of $N$(Dom($f$)$\cup\{u\}$))= the induced subgraph of $N$(Ran($f$)$\cup\{v\}$) holds. If it does not hold, abandon the vertex $v$ and go to **Step 6**, considering another "unvisited" vertex $v$ in graph $G_2$; Otherwise, go to **Step 13**;

**Step 13.** Let $G_1 = G'_1$, $G_2 = G'_2$, add the vertex pair $(u, v)$ to $f$, namely $v = f(u)$, clear up all "visited" tag in $G_2$, go to **Step 3**;

**Step 14.** End of While;

**Step 15.** Return the isomorphic function $f$ and stop the algorithm;

The **Step 11** of the subdivision algorithm is to introduce the "induced subgraph + remaining subgraph" strategy for subdividing a partition determined by a certain invariant $\zeta$. However, it is still the calculation of invariant $\zeta$. In **Step 12**, we determine whether $f$(the induced subgraph of $N$(Dom($f$)$\cup\{u\}$))= the induced subgraph of $N$(Ran($f$)$\cup\{v\}$) holds. This strategy can directly build the reconstruction from the subgraph to the parent graph. Its filtering ability is stronger, which can further reduce the number of backtracking and improve the overall efficiency of the algorithm.

## 5 Case Analysis

The following is an analysis of several typical cases (simple undirected graphs) to verify the correctness of our verification method and subdivision method.

### 5.1 Case Analysis of Verification Method

The judgment condition $\zeta$ is the characteristic polynomial of the distance matrix[5]. Reference [5] uses the relevant theories of linear algebra and matrix theory to "prove" that the characteristic polynomial of the distance matrix is a sufficient and necessary condition. This section proves through a positive case (**Example 1**) and a negative case (**Example 2**) that the "characteristic polynomial of the distance matrix" is not a necessary and sufficient condition.

**Example 1**. The distance matrices of the two undirected graphs shown in Figure 1 are shown in Figure 1(b).

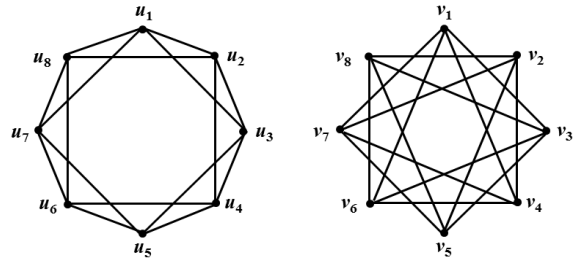

(a) The original graphs $G_1$ 和 $G_2$

(b) The distance matrices $D_1$ 和 $D_2$

Figure 1. Undirected graph with $n=8$ (isomorphic)

The eigenvalues calculated based on the distance matrices $D_1$ and $D_2$ are -3.41421, -3.41421, -2, -0.585786, -0.585786, 0, 0, 10. So $\zeta(D_1)=\zeta(D_2)$.

**Step 1.** Consider the vertex $u_1$ in $G_1$ and the vertex $v_1$ in $G_2$. The eigenvalues of $|\lambda I - D_{G_1 - u_1}|$ and $|\lambda I - D_{G_2 - v_1}|$ are -3.41421, -2.7574, -1.56801, -0.585786, -0.269111, 0, 8.59452. So $|\lambda I - D_{G_1 - u_1}| = |\lambda I - D_{G_2 - v_1}|$, $u_1 \leftrightarrow v_1$, adds the vertex pairs ($u_1$, $v_1$) to $f$, namely $v_1 = f(u_1)$.

**Step 2.** Consider the vertex $u_2$ in $G_1 - u_1$. The eigenvalues of $|\lambda I - D_{G_1 - u_1 - u_2}|$ are -3.87939, -1.6527, -1.17762, -0.467911, -0.338922, 7.51654. Consider the vertex $v_3$ in $G_2 - v_1$. The eigenvalues of $|\lambda I - D_{G_2 - v_1 - v_3}|$ are -2.83362, -2.61803, -1.20357, -0.381966, 0, 7.03719. Therefore, $|\lambda I - D_{G_1 - u_1 - u_2}| \neq |\lambda I - D_{G_2 - v_1 - v_3}|$, abandons $v_3$.

Consider the vertex $v_4$ in $G_2 - v_1$. $|\lambda I - D_{G_1 - u_1 - u_2}| = |\lambda I - D_{G_2 - v_1 - v_4}|$, so $u_2 \leftrightarrow v_4$. Further, $(u_1, u_2) \Leftrightarrow (f(u_1), f(u_2)) \Leftrightarrow (v_1, v_4)$, adds the vertex pairs ($u_2, v_4$) to $f$, namely $v_4 = f(u_2)$.

**Step 3.** Consider the vertex $u_3$ in $G_1 - u_1 - u_2$. The eigenvalues of $|\lambda I - D_{G_1 - u_1 - u_2 - u_3}|$ are -2.61803, -1.57093, -0.721651, -0.381966, 5.29258. Consider the vertex $v_3$ in $G_2 - v_1 - v_4$. The eigenvalues of $|\lambda I - D_{G_2 - v_1 - v_4 - v_3}|$ are -3.5669, -1.19707, -1, -0.452118, 6.21609. Therefore, $|\lambda I - D_{G_1 - u_1 - u_2 - u_3}| \neq$

$|\lambda I\text{-}D_{G_2-v_1-v_4-v_3}|$, abandons $v_3$.

Consider the vertex $v_6$ in $G_2\text{-}v_1\text{-}v_4$. $|\lambda I\text{-}D_{G_1-u_1-u_2-u_3}|=|\lambda I\text{-}D_{G_2-v_1-v_4-v_6}|$, so $u_3 \leftrightarrow v_6$. Further, $(u_1, u_2) \Leftrightarrow (f(u_1), f(u_2)) \Leftrightarrow (v_1, v_4)$, $(u_1, u_3) \Leftrightarrow (f(u_1), f(u_3)) \Leftrightarrow (v_1, v_6)$, $(u_2, u_3) \Leftrightarrow (f(u_2), f(u_3)) \Leftrightarrow (v_4, v_6)$, adds the vertex pairs $(u_3, v_6)$ to $f$, namely $v_6 = f(u_3)$.

**Step 4.** Consider the vertex $u_4$ in $G_1\text{-}u_1\text{-}u_2\text{-}u_3$. The eigenvalues of $|\lambda I\text{-}D_{G_1-u_1-u_2-u_3-u_4}|$ are -2, -1, -0.561553, 3.56155. Consider the vertex $v_2$ in $G_2\text{-}v_1\text{-}v_4\text{-}v_6$. The eigenvalues of $|\lambda I\text{-}D_{G_2-v_1-v_4-v_6-v_2}|$ are -2.38318, -1, -0.716463, 4.09965. Therefore, $|\lambda I\text{-}D_{G_1-u_1-u_2-u_3-u_4}| \neq |\lambda I\text{-}D_{G_2-v_1-v_4-v_6-v_2}|$, abandons $v_2$.

Similarly, the eigenvalues of $|\lambda I\text{-}D_{G_2-v_1-v_4-v_6-v_5}|$ are -3.41421, -1.16228, -0.585786, 5.16228. The eigenvalues of $|\lambda I\text{-}D_{G_2-v_1-v_4-v_6-v_8}|$ are -2.38318, -1, -0.716463, 4.09965. Therefore, $|\lambda I\text{-}D_{G_1-u_1-u_2-u_3-u_4}| \neq |\lambda I\text{-}D_{G_2-v_1-v_4-v_6-v_5}|$, abandons $v_5$. And $|\lambda I\text{-}D_{G_1-u_1-u_2-u_3-u_4}| \neq |\lambda I\text{-}D_{G_2-v_1-v_4-v_6-v_8}|$, abandons $v_8$.

On the other hand, $|\lambda I\text{-}D_{G_1-u_1-u_2-u_3-u_4}|=|\lambda I\text{-}D_{G_2-v_1-v_4-v_6-v_3}|$. $(u_1, u_4)$ is not an edge of $G_1$, but $(f(u_1), f(u_4)) \Leftrightarrow (v_1, v_3)$ is an edge of $G_2$. Not satisfy the constraints of **Step 9**, abandons $v_3$.

Similarly, $|\lambda I\text{-}D_{G_1-u_1-u_2-u_3-u_4}|=|\lambda I\text{-}D_{G_2-v_1-v_4-v_6-v_7}|$. $(u_1, u_4)$ is not an edge of $G_1$, but $(f(u_1), f(u_4)) \Leftrightarrow (v_1, v_7)$ is an edge of $G_2$. Not satisfy the constraints of **Step 9**, abandons $v_7$.

Thus, there is no corresponding point in $G_2\text{-}v_1\text{-}v_4\text{-}v_6$ that satisfies the "isomorphism definition (**Step 9**)" relationship for the vertex $u_4$ in $G_1\text{-}u_1\text{-}u_2\text{-}u_3$. So the judgment condition $\zeta$ "the characteristic polynomial of the distance matrix" is not a sufficient and necessary condition.

In fact, the isomorphic function $f$ in this example is $\{u_1 \leftrightarrow v_1, u_2 \leftrightarrow v_4, u_3 \leftrightarrow v_7, u_4 \leftrightarrow v_2, u_5 \leftrightarrow v_5, u_6 \leftrightarrow v_8, u_7 \leftrightarrow v_3, u_8 \leftrightarrow v_6\}$.

**Example 2.** In 1973, Paulus [14] made a complete enumeration of the conference matrices of order 25. He finds that up to isomorphism there are 15 strongly regular graphs with parameters $v = 25$, $k = 12$, $\lambda = 5$, $\mu = 6$ (and spectrum $12^1 2^{12}(-3)^{12}$).

Table 1. Some properties of 15 conference graphs on 25 vertices

| name | group size | two-graph | complement | max cliques | comments |
|---|---|---|---|---|---|
| P25. 01 | 1 | A | P25. 02 | $3^7, 4^{74}, 5^3$ | |
| P25. 02 | 1 | A | P25. 01 | $3^5, 4^{74}, 5^3$ | |
| P25. 03 | 2 | A | P25. 04 | $3^8, 4^{72}, 5^3$ | |
| P25. 04 | 2 | A | P25. 03 | $3^8, 4^{72}, 5^3$ | |
| P25. 05 | 2 | A | P25. 06 | $3^4, 4^{74}, 5^3$ | |
| P25. 06 | 2 | A | P25. 05 | $3^8, 4^{74}, 5^3$ | |
| P25. 07 | 6 | A | P25. 08 | $3^{14}, 4^{68}, 5^3$ | |
| P25. 08 | 6 | A | P25. 07 | $3^{14}, 4^{68}, 5^3$ | |
| P25. 09 | 6 | B | P25. 10 | $3^{54}, 4^{58}, 5^3$ | |
| P25. 10 | 6 | B | P25. 09 | $3^{54}, 4^{58}, 5^3$ | |
| P25. 11 | 72 | B | P25. 12 | $3^{36}, 4^{64}, 5^3$ | |
| P25. 12 | 72 | B | P25. 11 | $3^{84}, 4^4, 5^{15}$ | LS(5) |
| P25. 13 | 3 | C | P25. 14 | $3^3, 4^{75}, 5^3$ | |
| P25.14 | 3 | C | P25.13 | $3^1, 4^{75}, 5^3$ | |
| P25.15 | 600 | D | P25.15 | $3^{100}, 5^{15}$ | Paley(25) |

The vertex numbers associated with each vertex are given below, using it easy to generate the corresponding adjacency matrix.

Table 2. The vertex numbers of graphs P25.01 and P25.02

| P25.01 | P25.02 |
|---|---|
| 1, 2, 3, 4, 5, 6, 7, 8, 9, 10, 11, 12; | 1, 2, 3, 4, 5, 6, 7, 8, 9, 10, 11, 12; |
| 0, 2, 3, 4, 5, 6, 13, 14, 15, 16, 17, 18; | 0, 2, 3, 4, 5, 6, 13, 14, 15, 16, 17, 18; |
| 0, 1, 3, 4, 5, 6, 19, 20, 21, 22, 23, 24; | 0, 1, 3, 4, 5, 6, 19, 20, 21, 22, 23, 24; |
| 0, 1, 2, 7, 8, 9, 13, 14, 15, 19, 20, 21; | 0, 1, 2, 7, 8, 9, 13, 14, 15, 19, 20, 21; |
| 0, 1, 2, 7, 10, 11, 13, 16, 17, 19, 22, 23; | 0, 1, 2, 7, 10, 11, 13, 16, 17, 19, 22, 23; |
| 0, 1, 2, 8, 10, 12, 14, 16, 18, 20, 22, 24; | 0, 1, 2, 8, 10, 12, 14, 16, 18, 20, 22, 24; |
| 0, 1, 2, 9, 11, 12, 15, 17, 18, 21, 23, 24; | 0, 1, 2, 9, 11, 12, 15, 17, 18, 21, 23, 24; |
| 0, 3, 4, 8, 9, 10, 13, 15, 18, 22, 23, 24; | 0, 3, 4, 8, 9, 10, 13, 15, 18, 22, 23, 24; |
| 0, 3, 5, 7, 9, 11, 14, 16, 17, 21, 22, 24; | 0, 3, 5, 7, 9, 12, 14, 16, 17, 21, 22, 23; |
| 0, 3, 6, 7, 8, 12, 16, 17, 18, 19, 20, 23; | 0, 3, 6, 7, 8, 11, 16, 17, 18, 19, 20, 24; |
| 0, 4, 5, 7, 11, 12, 13, 14, 18, 20, 21, 23; | 0, 4, 5, 7, 11, 12, 13, 14, 18, 19, 21, 24; |
| 0, 4, 6, 8, 10, 12, 14, 15, 17, 19, 21, 22; | 0, 4, 6, 9, 10, 12, 14, 15, 16, 19, 20, 23; |
| 0, 5, 6, 9, 10, 11, 13, 15, 16, 19, 20, 24; | 0, 5, 6, 8, 10, 11, 13, 15, 17, 20, 21, 22; |
| 1, 3, 4, 7, 10, 12, 15, 16, 17, 20, 21, 24; | 1, 3, 4, 7, 10, 12, 15, 16, 17, 20, 21, 24; |
| 1, 3, 5, 8, 10, 11, 15, 16, 18, 19, 21, 23; | 1, 3, 5, 8, 10, 11, 15, 16, 18, 19, 21, 23; |
| 1, 3, 6, 7, 11, 12, 13, 14, 18, 19, 22, 24; | 1, 3, 6, 7, 11, 12, 13, 14, 18, 20, 22, 23; |
| 1, 4, 5, 8, 9, 12, 13, 14, 17, 19, 23, 24; | 1, 4, 5, 8, 9, 11, 13, 14, 17, 20, 23, 24; |
| 1, 4, 6, 8, 9, 11, 13, 16, 18, 20, 21, 22; | 1, 4, 6, 8, 9, 12, 13, 16, 18, 19, 21, 22; |
| 1, 5, 6, 7, 9, 10, 14, 15, 17, 20, 22, 23; | 1, 5, 6, 7, 9, 10, 14, 15, 17, 19, 22, 24; |
| 2, 3, 4, 9, 11, 12, 14, 15, 16, 20, 22, 23; | 2, 3, 4, 9, 10, 11, 14, 17, 18, 20, 21, 22; |
| 2, 3, 5, 9, 10, 12, 13, 17, 18, 19, 21, 22; | 2, 3, 5, 9, 11, 12, 13, 15, 16, 19, 22, 24; |
| 2, 3, 6, 8, 10, 11, 13, 14, 17, 20, 23, 24; | 2, 3, 6, 8, 10, 12, 13, 14, 17, 19, 23, 24; |
| 2, 4, 5, 7, 8, 11, 15, 17, 18, 19, 20, 24; | 2, 4, 5, 7, 8, 12, 15, 17, 18, 19, 20, 23; |
| 2, 4, 6, 7, 9, 10, 14, 16, 18, 19, 21, 24; | 2, 4, 6, 7, 8, 11, 14, 15, 16, 21, 22, 24; |
| 2, 5, 6, 7, 8, 12, 13, 15, 16, 21, 22, 23. | 2, 5, 6, 7, 9, 10, 13, 16, 18, 20, 21, 23. |

Table 3. The vertex numbers of graphs P25.03 and P25.04

| P25.03 | P25.04 |
|---|---|
| 1, 2, 3, 4, 5, 6, 7, 8, 9, 10, 11, 12; | 1, 2, 3, 4, 5, 6, 7, 8, 9, 10, 11, 12; |
| 0, 2, 3, 4, 5, 6, 13, 14, 15, 16, 17, 18; | 0, 2, 3, 4, 5, 6, 13, 14, 15, 16, 17, 18; |
| 0, 1, 3, 4, 5, 6, 19, 20, 21, 22, 23, 24; | 0, 1, 3, 4, 5, 6, 19, 20, 21, 22, 23, 24; |
| 0, 1, 2, 7, 8, 9, 13, 14, 15, 19, 20, 21; | 0, 1, 2, 7, 8, 9, 13, 14, 15, 19, 20, 21; |
| 0, 1, 2, 7, 10, 11, 13, 16, 17, 19, 22, 23; | 0, 1, 2, 7, 10, 11, 13, 16, 17, 19, 22, 23; |
| 0, 1, 2, 8, 10, 12, 14, 16, 18, 20, 22, 24; | 0, 1, 2, 8, 10, 12, 14, 16, 18, 20, 22, 24; |
| 0, 1, 2, 9, 11, 12, 15, 17, 18, 21, 23, 24; | 0, 1, 2, 9, 11, 12, 15, 17, 18, 21, 23, 24; |
| 0, 3, 4, 8, 9, 10, 13, 15, 18, 22, 23, 24; | 0, 3, 4, 8, 9, 10, 13, 16, 18, 21, 23, 24; |
| 0, 3, 5, 7, 9, 11, 14, 16, 17, 20, 23, 24; | 0, 3, 5, 7, 9, 12, 15, 16, 17, 19, 22, 24; |
| 0, 3, 6, 7, 8, 12, 16, 17, 18, 19, 21, 22; | 0, 3, 6, 7, 8, 11, 14, 17, 18, 20, 22, 23; |
| 0, 4, 5, 7, 11, 12, 13, 14, 18, 19, 21, 24; | 0, 4, 5, 7, 11, 12, 13, 14, 15, 20, 23, 24; |
| 0, 4, 6, 8, 10, 12, 14, 15, 17, 19, 20, 23; | 0, 4, 6, 9, 10, 12, 14, 16, 17, 19, 20, 21; |
| 0, 5, 6, 9, 10, 11, 13, 15, 16, 20, 21, 22; | 0, 5, 6, 8, 10, 11, 13, 15, 18, 19, 21, 22; |
| 1, 3, 4, 7, 10, 12, 15, 16, 17, 20, 21, 24; | 1, 3, 4, 7, 10, 12, 15, 17, 18, 20, 21, 22; |
| 1, 3, 5, 8, 10, 11, 15, 16, 18, 19, 21, 23; | 1, 3, 5, 9, 10, 11, 15, 16, 18, 19, 20, 23; |
| 1, 3, 6, 7, 11, 12, 13, 14, 18, 20, 22, 23; | 1, 3, 6, 8, 10, 12, 13, 14, 17, 19, 23, 24; |
| 1, 4, 5, 8, 9, 12, 13, 14, 17, 21, 22, 23; | 1, 4, 5, 7, 8, 11, 14, 17, 18, 19, 21, 24; |
| 1, 4, 6, 8, 9, 11, 13, 16, 18, 19, 20, 24; | 1, 4, 6, 8, 9, 11, 13, 15, 16, 20, 22, 24; |
| 1, 5, 6, 7, 9, 10, 14, 15, 17, 19, 22, 24; | 1, 5, 6, 7, 9, 12, 13, 14, 16, 21, 22, 23; |
| 2, 3, 4, 9, 10, 11, 14, 17, 18, 20, 21, 22; | 2, 3, 4, 8, 11, 12, 14, 15, 16, 21, 22, 23; |
| 2, 3, 5, 8, 11, 12, 13, 15, 17, 19, 22, 24; | 2, 3, 5, 9, 10, 11, 13, 14, 17, 21, 22, 24; |
| 2, 3, 6, 9, 10, 12, 13, 14, 16, 19, 23, 24; | 2, 3, 6, 7, 11, 12, 13, 16, 18, 19, 20, 24; |
| 2, 4, 5, 7, 9, 12, 15, 16, 18, 19, 20, 23; | 2, 4, 5, 8, 9, 12, 13, 17, 18, 19, 20, 23; |
| 2, 4, 6, 7, 8, 11, 14, 15, 16, 21, 22, 24; | 2, 4, 6, 7, 9, 10, 14, 15, 18, 19, 22, 24; |
| 2, 5, 6, 7, 8, 10, 13, 17, 18, 20, 21, 23. | 2, 5, 6, 7, 8, 10, 15, 16, 17, 20, 21, 23. |

The eigenvalues generated by the distance matrix in graphs P25.01 to P25.04 are -4, -4, -4, -4, -4, -4, -4, -4, -4, -4, -4, -4, 1, 1, 1, 1, 1, 1, 1, 1, 1, 1, 1, 1, 36. So $|\lambda I\text{-}D_{G_1}|=|\lambda I\text{-}D_{G_2}|=|\lambda I\text{-}D_{G_3}|=|\lambda I\text{-}D_{G_4}|$. Here, $D_{G_i}$ is the corresponding distance matrices of P25.0$i$ (1≤$i$≤4). It is known that the graphs P25.01, P25.02, P25.03, and P25.04 are not isomorphic, so the "characteristic polynomial generated by the distance matrix" is not a sufficient and necessary condition for graph isomorphism. This example is a counterexample that directly proves that the characteristic polynomial of the distance matrix is not a necessary and sufficient condition. So that, the main conclusion in reference [5] is incorrect. In fact, analyzing the proof process of reference [5], it is not difficult to find that its proof method can also be applied to the "characteristic polynomial of the adjacency matrix", thus concluding that the "characteristic polynomial of the adjacency matrix" is a sufficient and necessary condition, which has been proven to be an incorrect conclusion by other researchers.

### 5.2 Case Analysis of Subdivision Method

This section discusses the subdivision comparison of initial partitions of several invariants through case studies. These invariants include vertex degree, degree sequence of adjacent vertices, XOR distance, characteristic polynomial of the distance matrix.

**Example 3.** Consider the tree shown in Figure 2.

Figure 2. Comparison case 1 of invariant subdivision

The partition result of the "vertex degree" invariant is $\{\{v_1, v_5, v_6, v_{12}, v_{13}\}, \{v_2, v_3, v_4, v_7,$

$v_8, v_9, v_{10}\}, \{v_{11}\}\}$.

The partition result of the "degree sequence of adjacent vertices of a vertex" invariant is $\{\{v_1, v_5, v_6\}, \{v_2, v_4, v_7\}, \{v_3, v_8, v_9\}, \{v_{10}\}, \{v_{11}\}, \{v_{12}, v_{13}\}\}$.

The partition result of the "characteristic polynomial of the distance matrix" invariant is $\{\{v_1, v_5\}, \{v_2, v_4\}, \{v_3\}, \{v_6\}, \{v_7\}, \{v_8\}, \{v_9\}, \{v_{10}\}, \{v_{11}\}, \{v_{12}, v_{13}\}\}$.

It is not difficult to find that the partition of the invariant "degree sequence of adjacent vertices" is finer than that of the invariant "degree of vertices", and the partition of the invariant "characteristic polynomial of distance matrix" is finer than that of the invariant "degree sequence of adjacent vertices of a vertex".

**Example 4.** Consider the simple undirected graph shown in Figure 3.

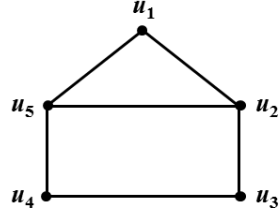

Figure 3.  Comparison case 2 of invariant subdivision

The partition result of the "vertex degree" invariant is $\{\{u_1, u_3, u_4\}, \{u_2, u_5\}\}$.

The partition result of the "degree sequence of adjacent vertices of a vertex" invariant is $\{\{u_1\}, \{u_2, u_5\}, \{u_3, u_4\}\}$.

The partition result of the "characteristic polynomial of the distance matrix" invariant is $\{\{u_1\}, \{u_2, u_5\}, \{u_3, u_4\}\}$.

From the view of automorphism groups, in the simple undirected graph shown in Figure 3, $u_2$ and $u_5$ can be isomorphic to each other, $u_3$ and $u_4$ can be isomorphic to each other. Therefore, the partition $\{\{u_1\}, \{u_2, u_5\}, \{u_3, u_4\}\}$ is the finest partition, i.e. optimal partition.

**Example 5.** Consider the P25.01 of the conference matrices of order 25.

The partition result of the "vertex degree" invariant is $\{\{u_1, u_2, u_3, u_4, u_5, u_6, u_7, u_8, u_9, u_{10}, u_{11}, u_{12}, u_{13}, u_{14}, u_{15}, u_{16}, u_{17}, u_{18}, u_{19}, u_{20}, u_{21}, u_{22}, u_{23}, u_{24}, u_{25}\}\}$.

The partition result of the "degree sequence of adjacent vertices of a vertex" invariant is $\{\{u_1, u_2, u_3, u_4, u_5, u_6, u_7, u_8, u_9, u_{10}, u_{11}, u_{12}, u_{13}, u_{14}, u_{15}, u_{16}, u_{17}, u_{18}, u_{19}, u_{20}, u_{21}, u_{22}, u_{23}, u_{24}, u_{25}\}\}$.

The partition result of the "characteristic polynomial of the distance matrix" invariant is $\{\{u_1, u_2, u_3, u_4, u_5, u_6, u_7, u_8, u_9, u_{10}, u_{11}, u_{12}, u_{13}, u_{14}, u_{15}, u_{16}, u_{17}, u_{18}, u_{19}, u_{20}, u_{21}, u_{22}, u_{23}, u_{24}, u_{25}\}\}$.

The partition result of the "characteristic polynomial of the distance matrix" invariant subdivided by induced subgraph + remaining subgraph is $\{\{u_1\}, \{u_2, u_3, u_{11}, u_{16}\}, \{u_4, u_7, u_{10}, u_{18}\}, \{u_5, u_6, u_{14}, u_{15}, u_{17}, u_{22}, u_{23}, u_{24}\}, \{u_8\}, \{u_9\}, \{u_{12}\}, \{u_{13}, u_{25}\}, \{u_{19}, u_{20}\}, \{u_{21}\}\}$.

It is not difficult to find that the partition of the invariant "degree sequence of adjacent vertices" is equal to that of the invariant "degree of vertices", and also is equal to that of the invariant "characteristic polynomial of distance matrix".

The specific data indicates the corresponding search space size, based on the initial partition obtained by using the invariant of the "characteristic polynomial of the distance matrix", is 25!= 15511210043330985984000000 (26 bits). The corresponding search space size, based on the partition obtained by using the induced subgraph + remaining subgraph for subdividing the initial partition of the "characteristic polynomial of the distance matrix" invariant, is 4!×4!×8!×2!×2! =92897280.

**Conclusion.** The initial partition of the invariant "degree sequence of adjacent vertices" is always finer than the initial partition of the invariant "degree sequence of vertices". The "induced subgraph + remaining subgraph" strategy can always subdivide the initial partition obtained from any necessary condition.

**Example 1 (continued).** Consider the two undirected graphs shown in Figure 4(a), whose distance matrices are shown in Figure 1(b). For the criterion of the characteristic polynomial of the distance matrix, by using the subdivision method of "induced subgraph+ remaining subgraph" strategy, our method can find isomorphic functions without backtracking. This is because our method of "induced subgraph + remaining subgraph" strategy has obtained partitions that are close to the optimal partition, thereby reducing the total number of iterations of backtracking.

The eigenvalues calculated based on the distance matrices $D_1$ and $D_2$ are -3.41421, -3.41421, -2, -0.585786, -0.585786, 0, 0, 10. So $\zeta(D_1)=\zeta(D_2)$.

**Step 1.** Consider the vertex $u_1$ in $G_1$ and the vertex $v_1$ in $G_2$. The eigenvalues of $|\lambda I-D_{G_1-u_1}|$ and $|\lambda I-D_{G_2-v_1}|$ are -3.41421, -2.7574, -1.56801, -0.585786, -0.269111, 0, 8.59452. So $|\lambda I-D_{G_1-u_1}|=|\lambda I-D_{G_2-v_1}|$, $u_1 \leftrightarrow v_1$, adds the vertex pairs $(u_1, v_1)$ to $f$, namely $v_1=f(u_1)$. The induced subgraphs of $N(u_1)$ and $N(v_1)$ are shown in Figure 4(b).

**Step 2.** Consider the vertex $u_2$ in $G_1$-$u_1$. The eigenvalues of $|\lambda I-D_{G_1-u_1-u_2}|$ are -3.87939, -1.6527, -1.17762, -0.467911, -0.338922, 7.51654. Consider the vertex $v_3$ in $G_2$-$v_1$. The eigenvalues of $|\lambda I-D_{G_2-v_1-v_3}|$ are -2.83362, -2.61803, -1.20357, -0.381966, 0, 7.03719. Therefore, $|\lambda I-D_{G_1-u_1-u_2}| \neq |\lambda I-D_{G_2-v_1-v_3}|$, abandons $v_3$.

Consider the vertex $v_4$ in $G_2$-$v_1$. $|\lambda I-D_{G_1-u_1-u_2}|=|\lambda I- D_{G_2-v_1-v_4}|$, so $u_2 \leftrightarrow v_4$. Further, $(u_1, u_2) \Leftrightarrow (f(u_1), f(u_2)) \Leftrightarrow (v_1, v_4)$. On the other hand, the induced subgraph of $N(\{u_1,u_2\})$ and the induced subgraph of $N(\{v_1,v_4\})$ are isomorphic, shown in Figure 4(c). As well as, the remaining subgraph of $G-N(\{u_1,u_2\})$ and the remaining subgraph of $G-N(\{v_1,v_4\})$ are isomorphic. Simultaneously, $u_1 \leftrightarrow v_1$ and $u_2 \leftrightarrow v_4$ are maintained. So adds the vertex pairs $(u_2, v_4)$ to $f$, namely $v_4=f(u_2)$.

**Step3.** $|\lambda I-D_{G_1-u_1-u_2-u_3}|=|\lambda I-D_{G_2-v_1-v_4-v_6}|$, we consider $u_3 \leftrightarrow v_6$. Further, $(u_1, u_2) \Leftrightarrow (f(u_1), f(u_2)) \Leftrightarrow (v_1, v_4)$, $(u_1, u_3) \Leftrightarrow (f(u_1), f(u_3)) \Leftrightarrow (v_1, v_6)$, $(u_2, u_3) \Leftrightarrow (f(u_2), f(u_3)) \Leftrightarrow (v_4, v_6)$. On the other hand, the induced subgraph of $N(\{u_1,u_2,u_3\})$ and the induced subgraph of $N(\{v_1,v_4,v_6\})$ are isomorphic, shown in Figure 4(d). As well as, the remaining subgraph of $G-N(\{u_1,u_2,u_3\})$ and the remaining subgraph of $G-N(\{v_1,v_4,v_6\})$ are isomorphic. But $u_1 \leftrightarrow v_1$ is not maintained. This means that the isomorphic function of the subgraph is not a part of the isomorphic function of the parent graph, so abandons $v_6$.

Similarly, $|\lambda I-D_{G_1-u_1-u_2-u_3}|=|\lambda I-D_{G_2-v_1-v_4-v_7}|$, we consider $u_3 \leftrightarrow v_7$. Further, $(u_1, u_2) \Leftrightarrow (f(u_1), f(u_2)) \Leftrightarrow (v_1, v_4)$, $(u_1, u_3) \Leftrightarrow (f(u_1), f(u_3)) \Leftrightarrow (v_1, v_7)$, $(u_2, u_3) \Leftrightarrow (f(u_2), f(u_3)) \Leftrightarrow (v_4, v_7)$. On the other hand, the induced subgraph of $N(\{u_1,u_2,u_3\})$ and the induced subgraph of $N(\{v_1,v_4,v_7\})$ are isomorphic, shown in Figure 4(e). As well as, the remaining subgraph of $G-N(\{u_1,u_2,u_3\})$ and the remaining subgraph of $G-N(\{v_1,v_4,v_7\})$ are isomorphic. Meanwhile, $u_1 \leftrightarrow v_1$, $u_2 \leftrightarrow v_4$, $u_3 \leftrightarrow v_7$ are maintained. This means that the isomorphic function of the subgraph is a part of the isomorphic function of the parent graph, so adds the vertex pairs $(u_3, v_7)$ to $f$, namely $v_7=f(u_3)$.

**Step4.** Finally, we find the isomorphic function $f=\{u_1\leftrightarrow v_1,\ u_2\leftrightarrow v_4,\ u_3\leftrightarrow v_7,\ u_4\leftrightarrow v_2,\ u_5\leftrightarrow v_5,\ u_6\leftrightarrow v_8,\ u_7\leftrightarrow v_3,\ u_8\leftrightarrow v_6\}$.

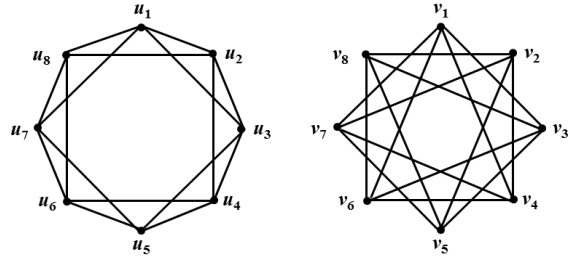

(a) The original graphs $G_1$和$G_2$

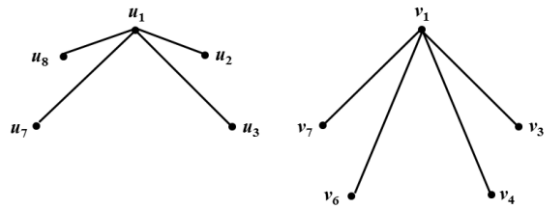

(b) The induced graphs $N(u_1)$ and $N(v_1)$

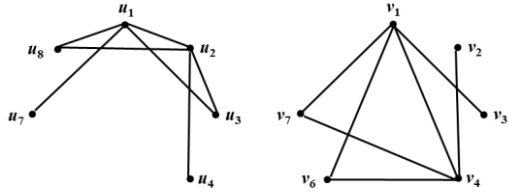

(c) The induced graphs $N(u_1,u_2)$ and $N(v_1,v_4)$

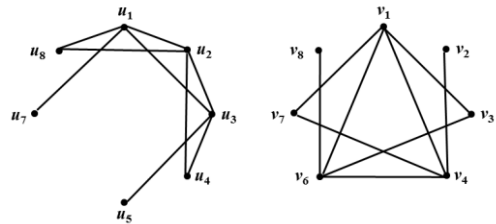

(d) The induced graphs $N(u_1,u_2,u_3)$ and $N(v_1,v_4,v_6)$

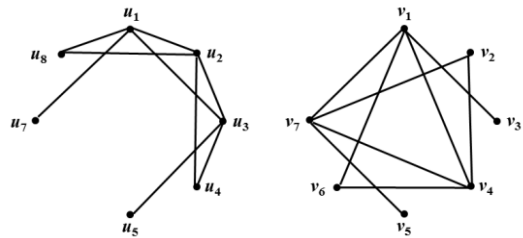

(e) The induced graphs $N(u_1,u_2,u_3)$ and $N(v_1,v_4,v_7)$

Figure 4. Undirected graph (isomorphic) with $n=8$ vertices and its subdivision process

# 6 Conclusion

All classifications, clustering, recognition, and learning problems related to the structure of a graph are related to the isomorphism determination problem of the graph. Up to now, the isomorphism determination problem of graphs still has important theoretical and practical value.

This article first proposes a verification method that can determine whether any judgment condition is sufficient and necessary in polynomial time. This verification method is independent of the proof process provided in the original literature that the judgment condition is a sufficient and necessary condition, so our method has universality. Secondly, a strategy of "induced subgraph + remaining subgraph" was proposed to further refine the initial partitioning obtained by necessary conditions. As the number of "suspected isomorphic corresponding vertices" is reduced as many as possible, this strategy can quickly narrow down the backtracking search space, thereby achieving high pruning efficiency. Regardless of whether the two graphs are isomorphic or not, the correct answer can be given in a shorter time. For two isomorphic graphs, an isomorphic function can be obtained.